\documentclass[journal]{IEEEtran}
\usepackage{graphicx}
\usepackage{epsf}
\usepackage{mathtools}
\usepackage{amssymb}
\usepackage{amsmath, amsfonts}
\usepackage{latexsym}
\usepackage{multirow}
\usepackage{multicol}
\usepackage[table]{xcolor}
\usepackage{color}

\usepackage{tabularx}
\usepackage{lipsum}

\usepackage{mathrsfs}

\usepackage{fixltx2e}

\usepackage[mathscr]{euscript}

\usepackage{commath}
\usepackage{MnSymbol}

\usepackage{mathtools} 
\DeclarePairedDelimiterX\setc[2]{[}{]}{\,#1 \;\delimsize\vert\; #2\,}
\DeclarePairedDelimiterX\parth[2]{(}{)}{\,#1 \;\delimsize\vert\; #2\,}

\PassOptionsToPackage{hyphens}{url}
\usepackage[unicode=true, bookmarks=true, bookmarksnumbered=true, bookmarksopen=true, bookmarksopenlevel=1, breaklinks=false, pdfborder={0 0 0}, pdfborderstyle={}, backref=false, colorlinks=false]{hyperref}

\usepackage[ruled,linesnumbered]{algorithm2e}

\usepackage{theorem}
{
\theorembodyfont{\rmfamily}

}

\usepackage{subcaption}
\usepackage{bbm}

\definecolor{orange}{RGB}{255,127,0}
\definecolor{blue}{RGB}{0,0,255}
\definecolor{red}{RGB}{255,0,0}
\definecolor{green}{RGB}{50,160,50}
\definecolor{grey}{RGB}{125,120,125}

\begin{document}

\title{\fontsize{24}{2}\selectfont Mitigation of Human RF Exposure in\\\vspace{0.1 in} Wearable Communications}
\author{Yakub Ahmed Sharif and Seungmo Kim, \textit{Member, IEEE}\\\{ys01063, seungmokim\}@georgiasouthern.edu\\Department of Electrical and Computer Engineering\\Georgia Southern University}
\maketitle

\begin{abstract}
A major concern regarding wearable communications is human
biological safety under exposure to radio frequency (RF) radiation generated by wearable devices. The biggest challenge in the implementation of wearable devices is to reduce the usage of energy to minimize the harmful impacts of exposure to RF on human health. Power management is one of the key energy-saving strategies used in wearable networks. Signals enter the receiver (Rx) from a transmitter (Tx) through the human body in the form of electromagnetic field (EMF) radiation produced during the transmission of the packet. It may have a negative effect on human health as a result of specific absorption rate (SAR). SAR is the amount of radio frequency energy consumed by human tissue in mass units. The higher the body's absorption rate, the more radio frequency radiation. Therefore, SAR can be reduced by distributing the power over a greater mass or tissue volume equivalently larger. The Institute of Electrical and Electronics Engineers (IEEE) 802.15.6-supported multi-hop topology is particularly useful for low-power embedded devices that can reduce consumption of energy by communicating to the receiver (Rx) through nearby transmitted devices. In this paper, we suggest a relaying mechanism to minimize the transmitted power and, as a consequence, the power density (PD), a measure of SAR.
\end{abstract}


\section{Introduction}
Wearable communications devices are rapidly making inroads, thanks to advancements in miniature electronics fabrication, wireless communications, batteries, and data analytics. The Wireless Body Area Network (WBAN) \cite{zigbee} is recently garnered for the autonomous application of wearable devices on the human body \cite{Astrin} for various applications from health monitoring to entertainment in real time. Despite the significant benefits that this technology will bring in, concerns related to human users' health have been rising.

Multi-hop communication was initially introduced for offloading the underlying network's traffic and thus enhancing the end-user performance of a network \cite{crowncom08}\cite{pimrc08}. Afterwards, the technology forms the basis of various advanced concepts such as ad-hoc networks \cite{crown11}\cite{apcc10} and distributed/decentralized networks \cite{globecom18}-\cite{bjkimaccess20}.

The concept can also be used in a WBAN. One key technological body is the IEEE 802.15.6. It supports a multi-hop extended star topology with the transmitter (Tx), the relay, and the Rx to enable short-range communication within the human body \cite{Higgins}\cite{Chen}. IEEE 802.15.6 recognizes the topologies of the star as well as an extended star topology. A relay node can be deployed for two main purposes: (i) when a direct link cannot be established or (ii) when there is a benefit via a relay (e.g., a lower energy consumption).

A major concern regarding wearable communications is human biological safety under exposure to electromagnetic field (EMF) generated by the wearable devices \cite{vtmag18}. A human body absorbs electromagnetic radiation via thermal or non-thermal absorption, which affect tissues on the absorbed area \cite{nonthermal}. Further, the EMF energy arriving at the human skin is dominated by the distance between the EMF emitting device and the skin \cite{nasim_cemag20}. This makes the wearable communications a type of wireless technologies on which the most careful analysis is needed regarding human EMF exposure.

\subsection{Contributions}
In this relation, this paper proposes a protocol reducing the power density to ensure a safe display of electromagnetic fields. The paper's contributions can be stressed as follows:
 
\begin{itemize}
\item First, while the prior works studied on adverse impacts of wearable communications due to exposure of RF only, this paper discusses the reduction of human RF exposure in the wearable communications.
\end{itemize}
\begin{itemize}
\item Second, using multi-hop communication in IEEE 802.15.6 standard the power density reduction by using a relaying method has been depicted. 
\end{itemize}
\begin{itemize}
\item Third, a new protocol has been proposed where the relay transmission power has been reduced to a certain extent so that the data rate ultimately becomes almost similar to that of direct connection without relay.
\end{itemize}

\section{Related Work}
Human EMF exposure in wireless communications garnered significant interest in recent literature. This section describes the state-of-the-art on characterization and reduction of EMF exposure in recent communications technologies.

\subsection{EMF Exposure in Recent Wireless Technologies}
Not only in wearable communications, other recent wireless technologies were also discussed on the human EMF exposure such as 5G \cite{secon19}-\cite{arxiv18}. The key finding from the prior works is that not only the uplink, but the downlink can also cause higher human EMF exposure compared to previous-generation. The main causes of such an increase in EMF exposure in 5G are (i) the adoption of larger sizes of a phased array and (ii) smaller cell sizes.

\subsection{Regulatory Efforts}
A wide range of applications for the use of electronic technologies and appliances (e.g. medical, military, athletics, etc.) are specifically related to the potentially adverse effects of concentrated tissue heating during RF radiation exposure. The fact that specific and strict guidelines are required in respect of human body exposure to RFs and standardization committees the Federal Communications Commission (FCC) and the International Commission on Non-Ionizing Radiation Protection (ICNIRP) identify SARs and set threshold levels in different parts of the body: (i) \textit{1.6 W/kg averaged over 1 g} of tissue for use against the head and 4.0 W/kg which is averaged over 10 g of tissue for use on the wrist. This cap is set for the USA, Canada, and South Korea \cite{4197534}; (ii) \textit{2 W/kg averaged over 10 g} of tissue for use against the head and 4.0 W/kg which is averaged over 10 g of tissue for use on the wrist, is recognized in the EU, Japan, and China \cite{icnirp}\cite{EMC EUROPE}.

\subsection{Multi-Hop Communications for EMF Reduction}
Several works discussed very specific aspects of relaying mechanism in wearable communications. One case study showed that implants such as pacemakers tend to relay tracking information to devices located outside the body \cite{kurup}. Relay nodes, if placed outside the human body, may provide stronger networks and lower energy limits. A relaying node re-selection problem is being addressed in the IEEE 802.15.6 Body Area Network (BAN) specification, which includes the relay node's energy consumption burden and optimizing the lifetime of the network \cite{Kim}. 

When the relay nodes are located in an ideal location for the sensor nodes to connect to the center through relay nodes, a strong performance rate of transmission of packet can be obtained and the negative effect of the SAR can be reduced by using particle swarm optimization (PSO) to determine the optimal location for the relay nodes in the body \cite{Ahmed6}. When the node produces normal data at the routing phase, relay transmission is used to find the best route, with the least hops \cite{Qu}. 

Another work uses the PSO algorithm to determine an optimal location for the relay node, in order to send packets via relay to the path with least SAR and thus to improve the packet transmission success rate \cite{Wu}. A network layer protocol is structured to handle the two-hop networking technologies and is responsible for the selection of relay and transmission of data \cite{Breva}. A relay-based improved throughput and energy-efficient multi-hop routing protocol (Rb-IEMRP) has been suggested there for the intra-wireless body sensor network (Intra-WBSN). In order to determine the total number of relay nodes to be used, statistical modeling was introduced. Emergency data is directly forwarded to body node coordinator (BNC) without the relay being used \cite{Rashid}. A feasible alternative for maintaining network access is the installation of a limited number of relay nodes, the key purpose of which is to interact with other sensors or relay nodes \cite{linn}. Some works suggested that with the help of the relay nodes the success rate of transmission and the network lifetime can be increased in WBAN. To analyze the human RF exposure on the internet of battlefield things one comprehensive analysis framework was provided where operating frequency is used as 60 GHz with a more directive antenna radiation pattern \cite{milcom19}\cite{kabir19}. An extensive review of SAR for different commercial wearable devices was provided where SAR comparison for the devices were shown at different carrier frequencies \cite{im4}. Here the current measurement methodologies for analyzing SAR were depicted along with the biological consequences from EMF exposure of wearable devices.

\section{System Model}

\subsection{Multi-Hop Networking}

\subsubsection{IEEE 802.15.6}
Scientists have long been criticizing SAR as insufficient as measured in a liquid-filled flat phantom \cite{Cleg}. This does not describe the complex features and interactions of the electromagnetic properties of living tissues, or RF signals. In terms of optimum achievement on the narrow band, all physical (PHY) overhead and medium access control (MAC) layers are taken into account. With the highest frequency and modulation order, the IEEE 802.15.6 standard can exceed 680 kbps \cite{Alam}. The IEEE 802.15.6 standard is used in this article, which covers data rates of up to 150 Mbps along with 802.11b. In addition to the transmission power change, different physical layer configurations are added to the IEEE 802.15.6 standard.

The specification under development should be supported by bit rates that are scalable according to the Task Group 6's findings, and the protocol is short in distance (0 to 5 m). It permits network sizes of 256 devices and guarantee extremely low latency and ultra-low consumption of power. The main aim of this design is to link low power embedded devices while promoting high rates of data and service quality. The specification defines two kinds of network topologies, in the extended star topology devices will exchange information through relay nodes. To allow the best range of transmission schemes, the carrier sense multiple access with collision avoidance (CSMA/CA) method is extended in IEEE 802.15.6 \cite{Ullah} to fit the best selection of the transmission scheme.

\subsubsection{Network Topology}
The framework is constructed using two-hop extended star topology to optimize the process of finding the relay node location. If the connection of the Tx with the Rx is low, or energy must be saved, the Tx will choose a relay facility.

The path from a Tx to a Rx is determined based on the link request and its allocation frame \cite{Lin}\cite{Maskooki}. A wake-up control system is used for a relay node re-selection \cite{Kim}. The human body is split into parts of various weights to understand the effects of SAR on the human body \cite{Wu}. The ideal location for the relay nodes is evaluated with the proposed algorithm. It also has the SAR as lowest and a higher distribution rate for packets.

\begin{figure}[t]
\centering
\includegraphics[width =\linewidth]{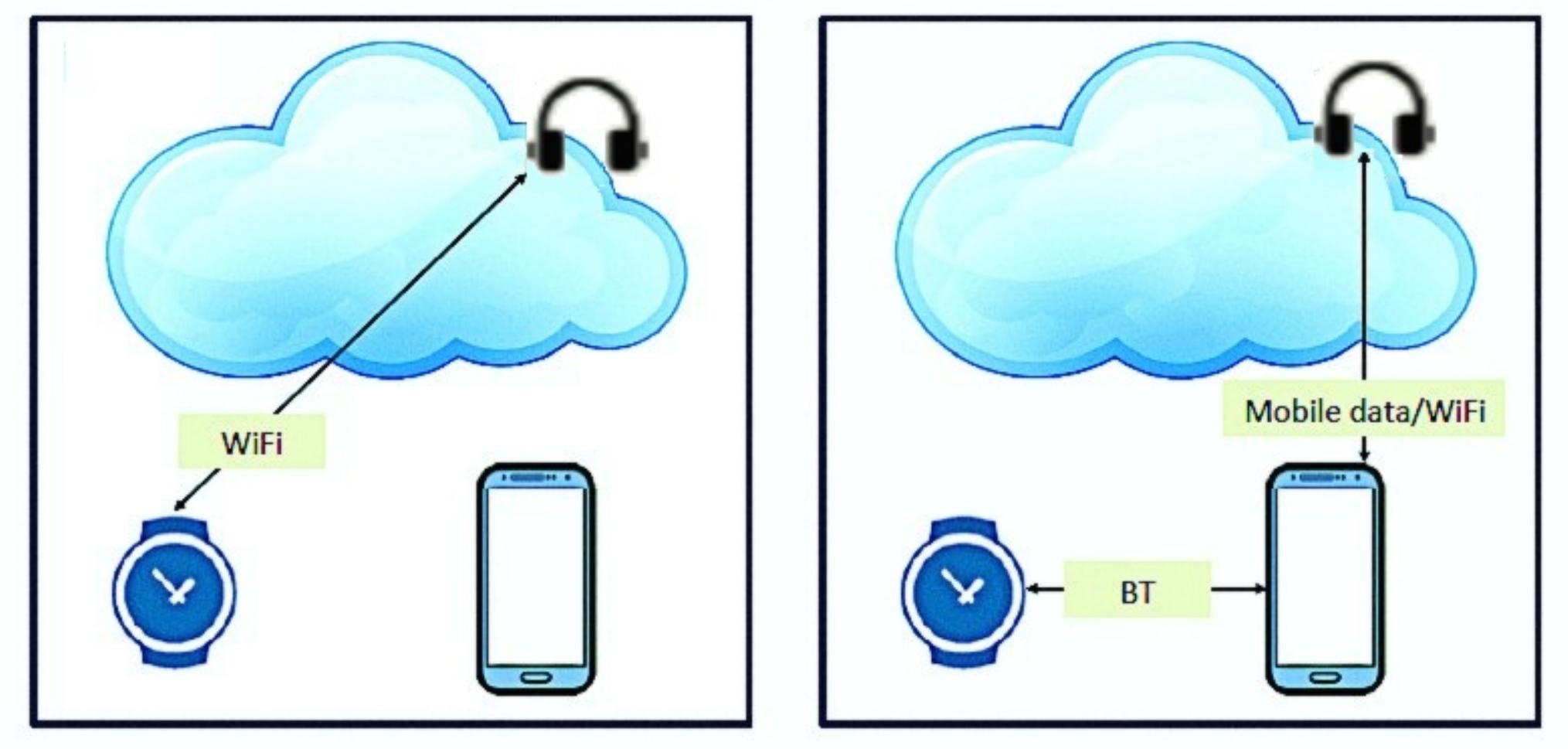}
\caption{The method of direct and multi-hop communication}
\label{fig_multihop}
\end{figure}

Relaying via a smartphone is considered more effective than direct communication over the internet. These internet communication modes are used by current smartwatch users. In one method, the smartwatch interacts directly with the mentioned servers, while in the other, it relays via paired Bluetooth of smartphone. In Fig. \ref{fig_multihop}, internet communication methods for smartwatch apps that use direct contact and relay through Bluetooth of cellphone have been shown. To minimize radiation, cell phone relays are built to automatically reduce power to the lowest possible amount in order to have a high-quality network. When used in places with strong coverage, a cell phone can function at a lower transmission power \cite{IOT}. A typical WBAN architecture used here involves a small network across the body (around 2 m) usually, the transceiver is the most power-consuming portion. To minimize power consumption each node must function autonomously. The multi-hop wireless networking methods increase node interdependence, augmenting consumption of power \cite{coll}\cite{wcnc16}.

\subsection{Human RF Exposure Measurement}
\subsubsection{PD}
Recent studies have demonstrated that PD is not as efficient as SAR or temperature when measuring exposure to EMF because SAR can display energy EMF levels, while PD cannot \cite{Collins}. PD is specified as the power radiation at a distance $d$ \cite{Rappaport} per unit volume, which is formally written as
\begin{equation}
\mathrm{PD}(d)=\frac{|E(d)|^{2}}{\rho_{0}}\left[\mathrm{W} / \mathrm{m}^{2}\right]
\end{equation}
where $\rho_{ 0}$ is the free space characteristic impedance and $E(d)$ is the RMS electric field. Notice that PD can be rewritten, using the transmission parameters, as
\begin{equation}
\operatorname{PD}(d, \phi)=\frac{P_{t} G_{t}(d, \phi)}{4 \pi d^{\alpha}}
\end{equation}
where $P_{t}$ is the transmitted power; $G_{t}$ is the transmit gain of antenna; $d$ is the distance (m) from the Tx to relay or Rx. $\alpha$ is the path loss exponent.

\subsubsection{SAR}
SAR is a function of PD and is characterized as a measurement of incident energy that is absorbed per unit of mass and time. As such, it is possible to quantify the amount of absorption rate from the electromagnetic field by the human body, which measures the power consumed per unit mass. A local SAR value at point m in W/kg is given by \cite{Hochwald}
\begin{equation}
\operatorname{SAR}(\mathrm{m})=\frac{\sigma|E(\mathrm{m})|^{2}}{\rho}[\mathrm{W} / \mathrm{kg}]
\end{equation}
where $\sigma$ is the material conductivity and $\rho$ gives the material density. The SAR value in terms of $d$ as a function of $\mathrm{PD}(d, \phi)$ for the wearable communications system can be characterized as \cite{Chahat}
\begin{equation}\label{eq_sar_dphi}
\operatorname{SAR}(d, \phi)=\frac{2\text{PD}(\phi) T(\phi) m(\phi)}{\delta \rho}
\end{equation}
where power transmission is defined as $T$, and $\delta$ is the skin penetration depth (m). The function $m(\phi) \cite{Hochwald}$ depends on the dielectric constant properties. The same Eq. (\ref{eq_sar_dphi}) at a point on the air-skin boundary \cite{Chahat} can be rewritten in the form of $R$ being the reflection coefficient \cite{Rappaport} as

\begin{equation}
\operatorname{SAR}(d, \phi)=\frac{2 \mathrm{PD}(d, \phi)\left(1-R^{2}\right)}{\delta \rho}
\end{equation}
where the mass density for tissue and the skin penetration depth are denoted by $\rho$ and $\delta$, respectively \cite{Collins}. Also, it is significant to notice that $d$ and $\phi$ are determined by the position of the EMF emitting device.

\subsection{RF Propagation}
The attenuation patterns of an antenna unit on the azimuth and elevation plane are given by \cite{dissertation}
\begin{equation}\begin{aligned}
&A_{a}(\phi)=\min \left\{12\left(\frac{\phi}{\phi_{3 d b}}\right)^{2}, A_{m}\right\}[\mathrm{dB}]\\
&A_{e}(\theta)=\min \left\{12\left(\frac{\theta-90^{\circ}}{\theta_{3 d b}}\right)^{2}, A_{m}\right\}[\mathrm{dB}]
\end{aligned}\end{equation}
where $\theta$ and $\phi$ are angles of a beam on the elevation and azimuth planes, respectively. Then the antenna unit pattern that is combined in the two planes is given by
\begin{equation}
A(\theta, \phi)=\min \left(A_{a}(\phi)+A_{e}(\theta), A_{m}\right)[\mathrm{dB}]
\end{equation}
where Am (= 30 dB) is a maximum attenuation (front-to-back ratio) \cite{Moorut}, but it can be higher in practice.
The general pattern equation is adopted which is given by \cite{lett17}
\begin{equation}
G(\theta)=G_{\max }-\exp (-2 \pi j \delta \sin \theta)\rm{~[dB]}
\end{equation}
where $\delta$ denotes the antenna unit separation distance, and $\theta$ denotes a general angle.
It is more desirable to assume a continuous line-of-sight (LOS) \cite{Heath} link between a wearable device and the human body. This paper adopts a propagation model for the calculation of a path loss (PL) given by
\begin{equation}
\mathrm{PL}=24 \log (d)+24 \log (f)+38.93\rm{~[dB]}
\end{equation}
where $f$ (GHz) represents the operating frequency and $d$ (m) is the separation distance between the antennas. The 2.4 GHz band has been studied as a primary frequency band for existing wearable systems.

\begin{figure}[t]
\centering
\includegraphics[width = \linewidth]{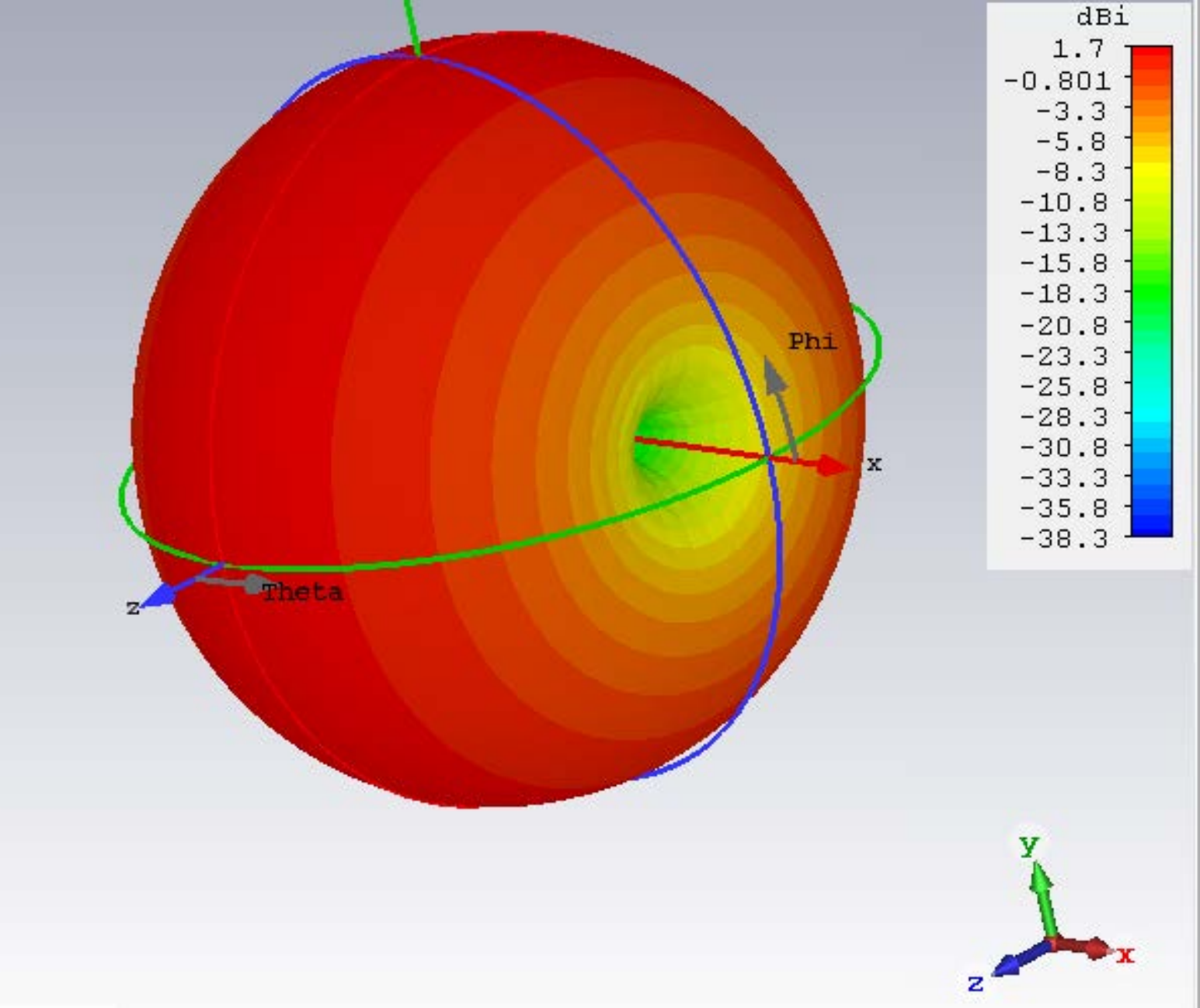}
\caption{Antenna radiation pattern in CST at 2.4 GHz}
\label{fig_cst}
\end{figure}

For wearable operations, different contact linkages require different patterns of radiation. When two on-body devices interact, the specifications seem to follow an omni-directional pattern. Nonetheless, a dipole pattern is more suitable when on-body devices interact with an outside-body system. The dipole pattern has been recorded \cite{Ahmed6} for support of these two operating states, in which an antenna can change its resonance to serve the same frequency on both states. This analysis suggests a dipole antenna in the pattern on a certain plane that is similar to omni-directional, as shown in Fig. \ref{fig_cst}.

\section{Proposed Protocol}

\subsection{Background}
The PD metric is used in the distance measurement so that it is used primarily for measuring far-field exposure in downlinks. In the downlink, the transmission power is diminished at least quadratically as long as the human body is far from access points. One easy way to reduce PD is to decrease transmission power or increase the spacing. A typical Tx power for a device that is portable is 23 dBm \cite{im4} (e.g., for Fitbit id XRAFB505 Tx power 15 dBm is used). Nonetheless, this power is not selected because there are no higher capacity, decent performance and long battery life amplifier. Rather, this choice is because some portable devices from various manufacturers are close to their SAR maximum with a power of 23 dBm, and more efficient Txs risk the system exceeding the limit. Therefore, the SAR restriction acts as a power limit in a single transmitting element system. We consider the highest potential exposure a human consumer can encounter. In other words, no technique for mitigation of received power is supposed in the system model that this article refers to.

\begin{algorithm}[t]
\SetAlgoLined
Input 1: $R$: Data rate without relay

Input 2: $\bar{R}$: Data rate with relay

Input 3: $D$: Distance from Tx to Rx

Input 4: $\bar{D}$': Distance from Tx to relay

\eIf{$D \leq \bar{D}$}{
\textbf{\%-- Distance --\%}\\
Total PD $\longleftarrow$ PD of single-hop
}
{
Total PD $\longleftarrow$ PD of multi-hop
}

\eIf{$R > \bar{R}$}{
\textbf{\%-- Data rate --\%}\\
Reduce the Tx power by a certain amount
}
{
No action required
}
\caption{A Class of Schemes for Power Control}
\label{algorithm_proposed}
\end{algorithm}

\begin{figure}[t]
\centering
\includegraphics[width =\linewidth]{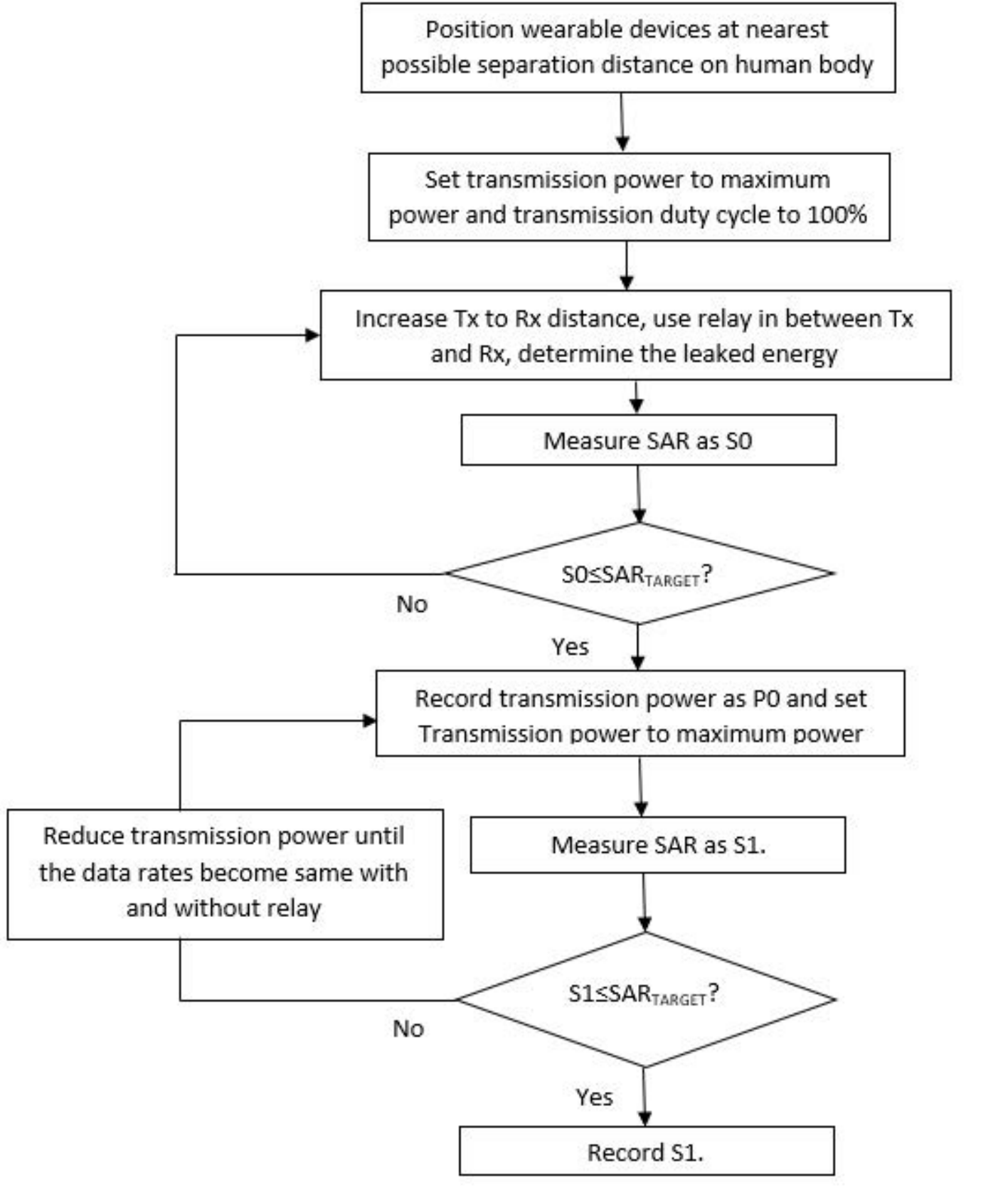}
\caption{Proposed protocol}
\label{fig_flowchart}
\end{figure}

\subsection{Human RF Exposure Reduction Protocol}
This paper now introduces a new protocol guaranteeing RF exposure below a threshold. Fig. \ref{fig_flowchart} provides functional information for the proposed protocol, where the wearable first deals with the maximum transmission power and then increase the distance to reduce SAR. The provided protocol also lets the system update the PD. The choice of the node is based on the input of the relayed node given by relays. SAR is measured when the device is operating at full transmit power in an expected manner. For a single-hop direct communication the Tx directly communicates with the Rx and the distance is measured. For multi-hop relay based communication, we place a relay station: the Tx communicates with the relay first and then the relay to the Rx communication is established. The distance and leaked energy are measured accordingly. Emergency data is directly delivered (single-hop) to the Rx. Normal data is delivered through multi-hop communication using the relay. From path loss received power is measured. From received power SNR and data rate are measured accordingly. SAR and PD are evaluated in terms of the heatmap in MATLAB simulation. If the threshold is exceeded for one metric EMF exposure is high. To mitigate, transmitted power is reduced to an extent so that the data rate with relay becomes the same as that of without relay. Algorithm \ref{algorithm_proposed} displays the scheme class and is defined by the following four parameters.

As the contact range of WBAN is 0 to 5 m, the whole body can be covered. Since the distance between nodes is small, the single-hop transmission is possible, but it impacts the body more with higher transmit power. In two-hop communication, the Tx interacts with the Rx using the relay node. The advantage of this transmission lies in the fact that the Tx can use less transmitting power \cite{Wu}.
SAR is determined when the system operates at maximum transmission power in an intended manner, as the electrical field is typically not uniform in space. Depending on the nature of the signal source, the short-term time average is achieved by the probe \cite{Hochald}.

\subsection{SAR control mechanisms}
One critical problem of a multi-hop topology is that there are multiple sources of RF emitting sources. In other words, there would be only one source of RF emission if it was a direct Tx-Rx link. However, with relay stations, although each of the Tx and the relay stations can operate at a lower Tx power, a \textit{aggregation} of the RF emissions may be significant enough to cause a health issue.

\begin{figure}[t]
\centering
\includegraphics[width =\linewidth]{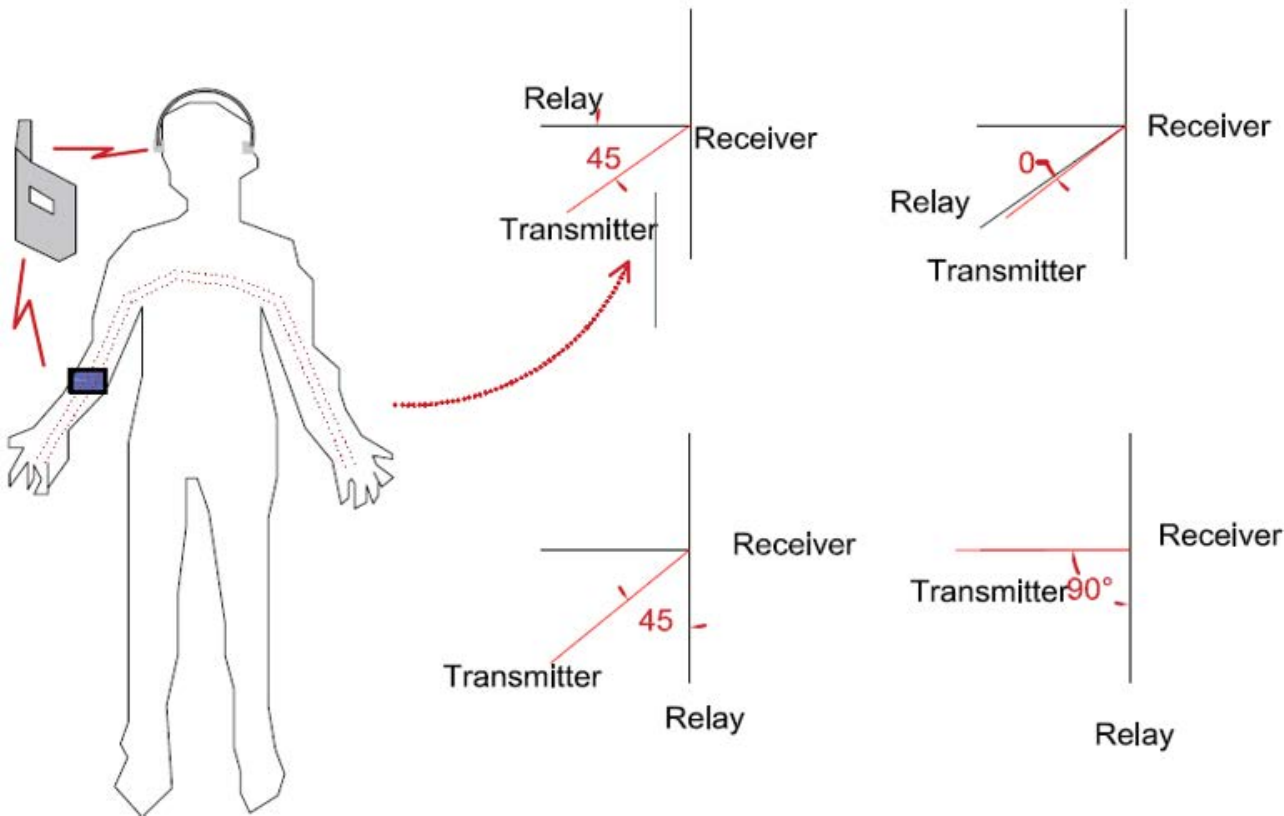}
\caption{Energy Leakage from Tx to Rx}
\label{fig_leakage}
\end{figure}

As such, we focus on such an aggregated energy in a multi-hop topology. Notice that we assume a beamforming technique for a Tx and a relay station \cite{patent}. We refer to the technique to measure the leaked energy determining the angle in between the Tx and the relay since some energy is leaked towards the Rx while communicating the Tx to the relay, as shown in Fig. \ref{fig_leakage}. This maximizes the received SNR while maintaining SAR compliance.

\begin{figure}[t]
\centering
\includegraphics[width =\linewidth]{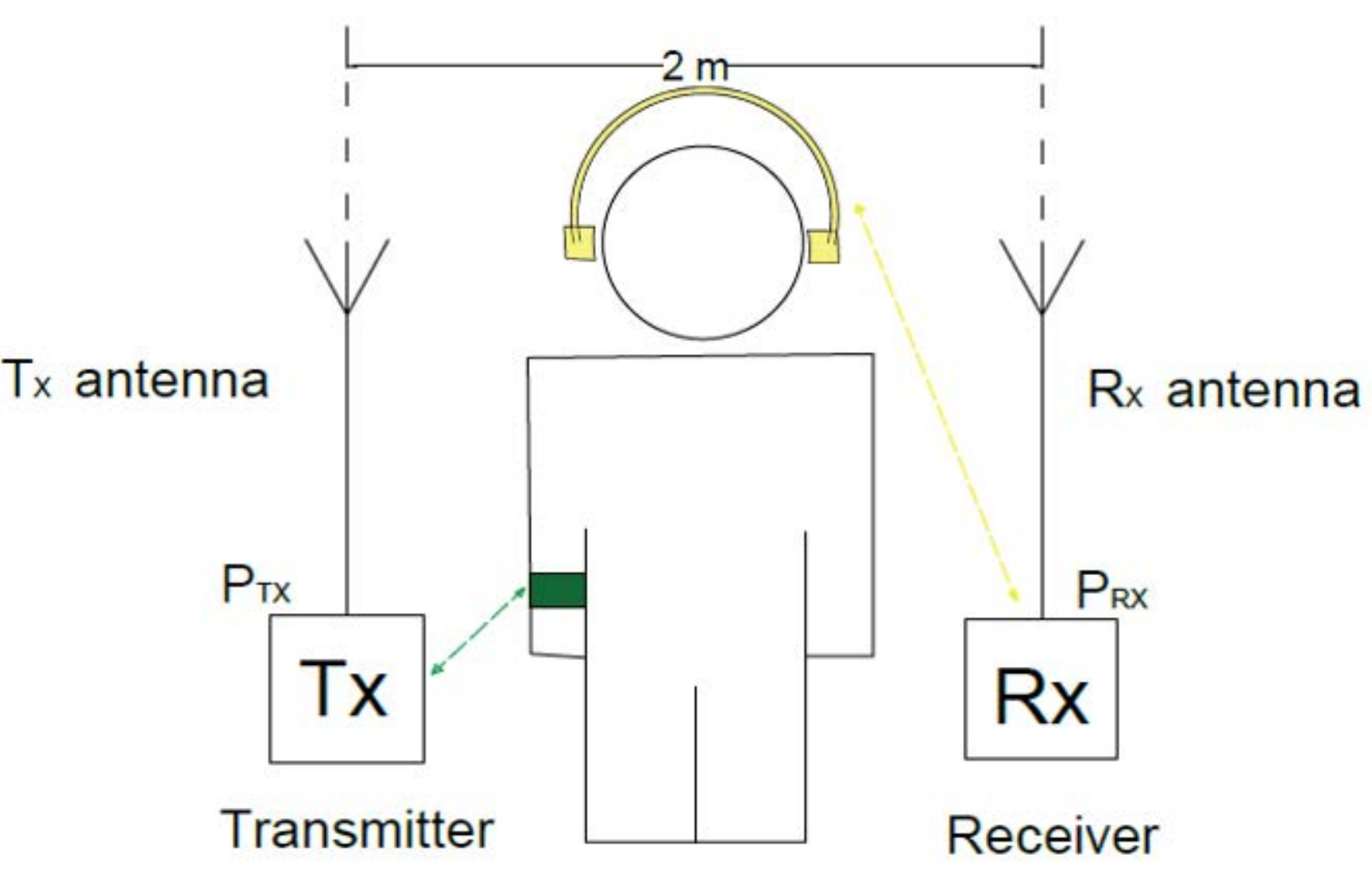}
\caption{Distance between the Tx and Rx antenna}
\label{fig_distance}
\end{figure}

The separation between the Tx and the Rx antenna has been shown in Fig. \ref{fig_distance}, where the Tx is searching for the closest device.

The operation of this routing protocol is completed in different phases. A time division duplex (TDD) specifies the transmission scheme to divide the channel into a time slot of 625 seconds, corresponding to different frequency hopping simulates full-duplex communication in the same transmission channel. The MAC layer must also adapt to this development. The IEEE 802.11 MAC layer has as its main difference, the mechanism enters the CSMA/CA as a method of medium access \cite{Oliveira}. The protocol stack follows the open systems interconnection (OSI) reference model on its PHY and MAC layers. The PHY layer has been developed to operate in the 2.4 GHz and 5 GHz, with DSSS and orthogonal frequency division multiplexing (OFDM).

When the focus is not the distance but the energy factor, its low power mode operation can also be used to reduce the device's transmission power. For example, in an 802.11 system, if the received signal strength indicator (RSSI) is large enough, power backoff can be used without adversely affecting throughput. Or, if the channel conditions are good enough, the highest power backoff can be chosen corresponding to the different overlapping areas, which will result in the lowest SAR value.

\section{Simulation Results}

\begin{table}[t]
\caption{Values for key parameters}
\centering
\begin{tabular}{ |l|l|}
\hline
\textbf{\cellcolor{gray!30}Parameter} & \cellcolor{gray!30}\textbf{Value}\\
\hline\hline
Bandwidth & 4 MHz \\
Temperature & 295 K\\
Noise figure & 19.2 dB\\
Grid length & 16 cm\\
Grid width & 15 cm\\
Skin permittivity & 39.2 \\
Skin conductivity & 1.8 S/m\\
Skin penetration depth & 113 mm\\
Tx power & 15 dBm\\
Relay Tx power & 15 dBm\\
Tx gain & 1.7 dBi\\
Relay Tx gain & 8 dBi\\
Antenna 3dB angle & 93 degrees\\
Maximum data rate & 10 Mbps\\
\hline
\end{tabular}
\label{table_parameters}
\end{table}

\subsection{Setting}
Simulations are carried out mostly in MATLAB and also using a 3D electromagnetic solver CSTMWS. For numerical analysis, simulation parameters are summarized in Table \ref{table_parameters}.

A simulation was done where the Tx (1,1) and Rx (15,15) were kept in a fixed position and the relay positions were varied in every possible coordinate using uniform distribution to check the distance and power density that fits the best. Without the relay, there remains a direct communication between the Tx and the Rx, where the distance remains constant between them, and PD depends upon that. Based on a great distance, PD can be lowered.

\subsection{Results and Discussions}
\subsubsection{Before Application of Proposed Protocol}
\begin{figure}[t]
\centering
\includegraphics[width = \linewidth]{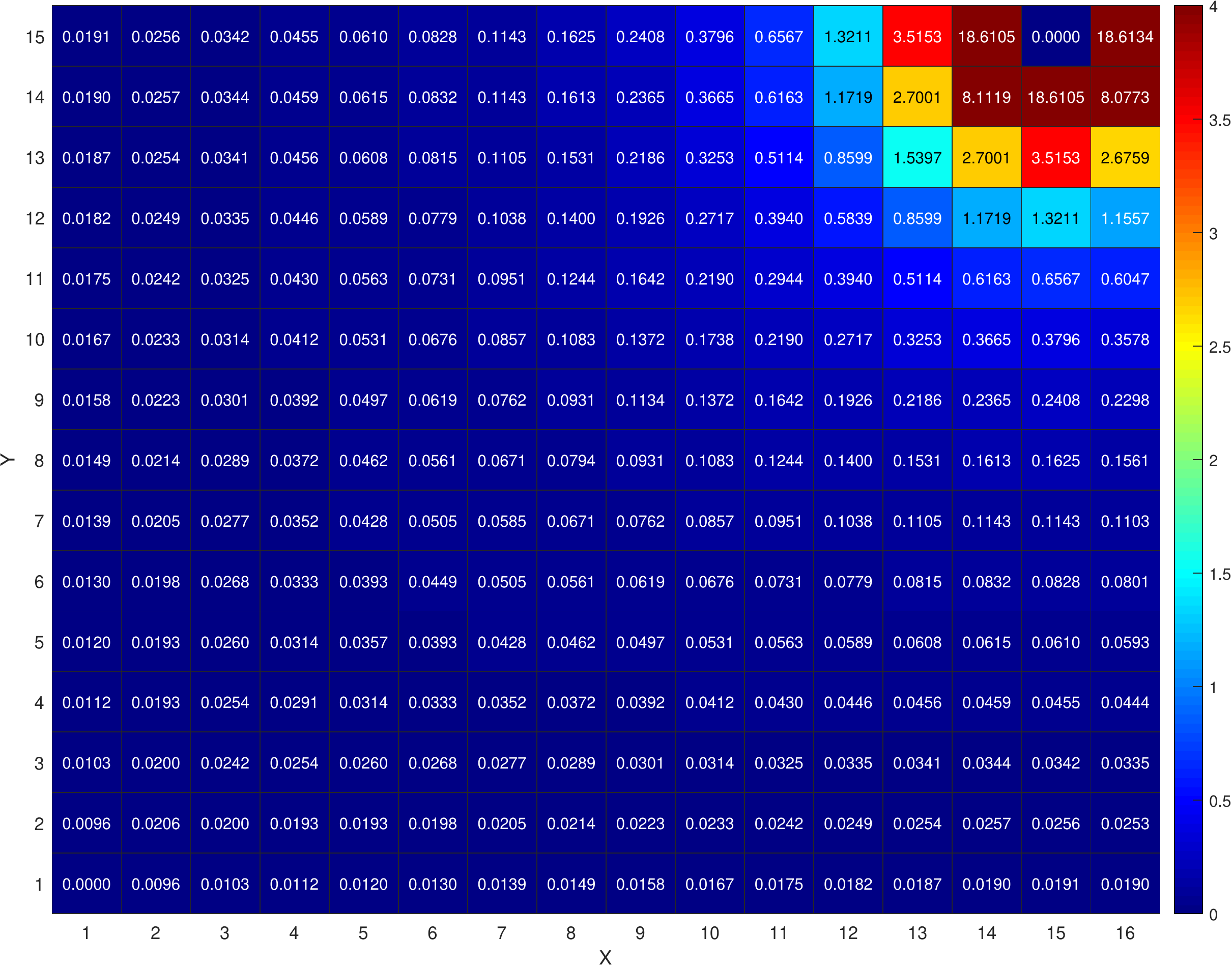}
\caption{SAR with the relay maximum near the Rx}
\label{fig_sar_before}
\vspace{0.2 in}
\centering
\includegraphics[width = \linewidth]{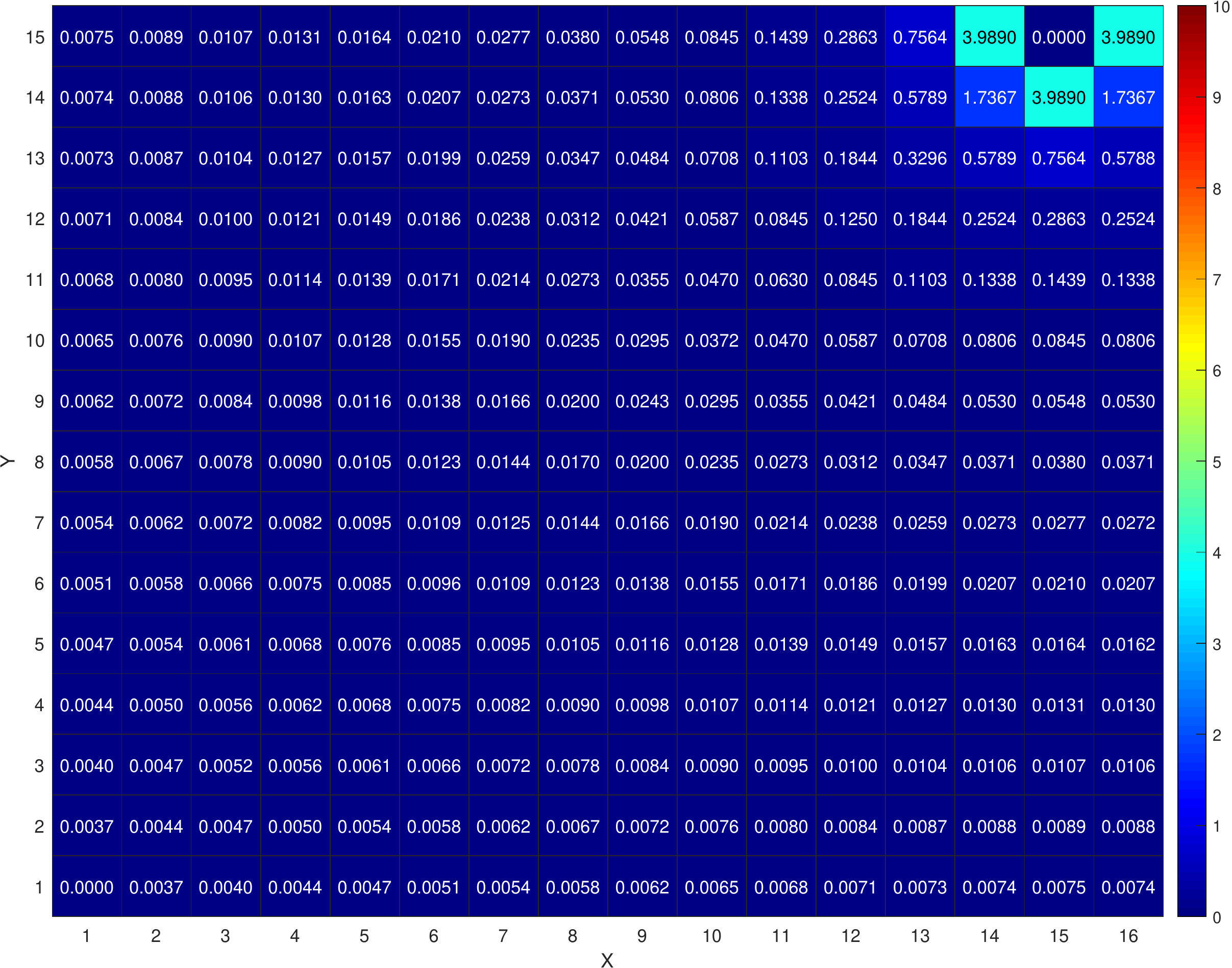}
\caption{Total PD with the relay maximum near the Rx}
\label{fig_pd_before}
\end{figure}

Figs. \ref{fig_sar_before} and \ref{fig_pd_before} show the heatmaps of SAR and power density for relay based communication. The heatmap is used to find the location of the relay node with the metrics surpassing the threshold. It was found that when the relay is near to the Rx the power density or SAR is the maximum.

\begin{figure}[t]
\centering
\includegraphics[width = \linewidth]{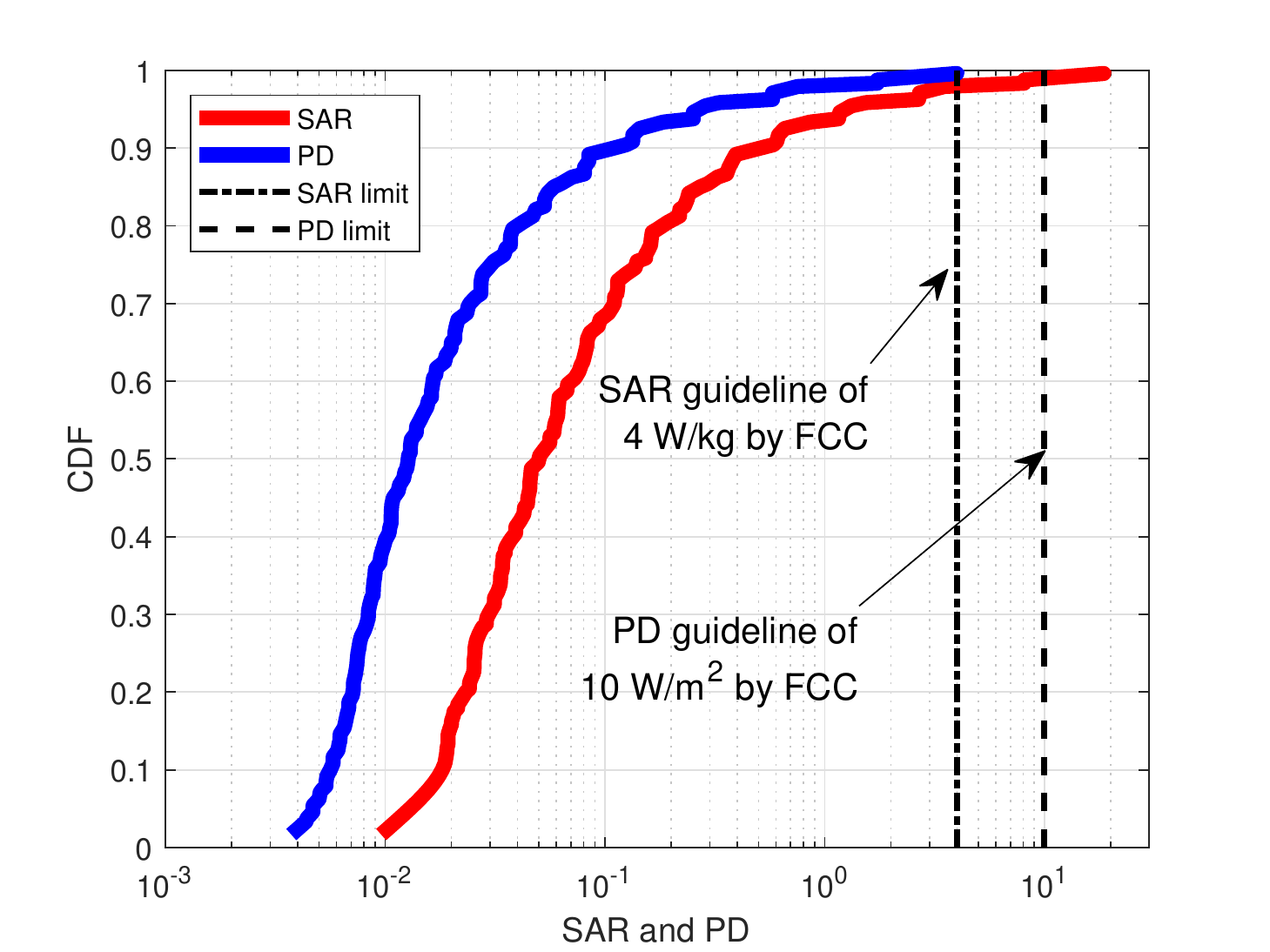}
\caption{CDF of SAR and PD with exposure limit}
\label{fig_cdf}
\end{figure}

A CDF of SAR and PD with the exposure limit is shown in Fig. \ref{fig_cdf}.

\subsubsection{After Application of Proposed Protocol}
\begin{figure}[t]
\centering
\includegraphics[width = \linewidth]{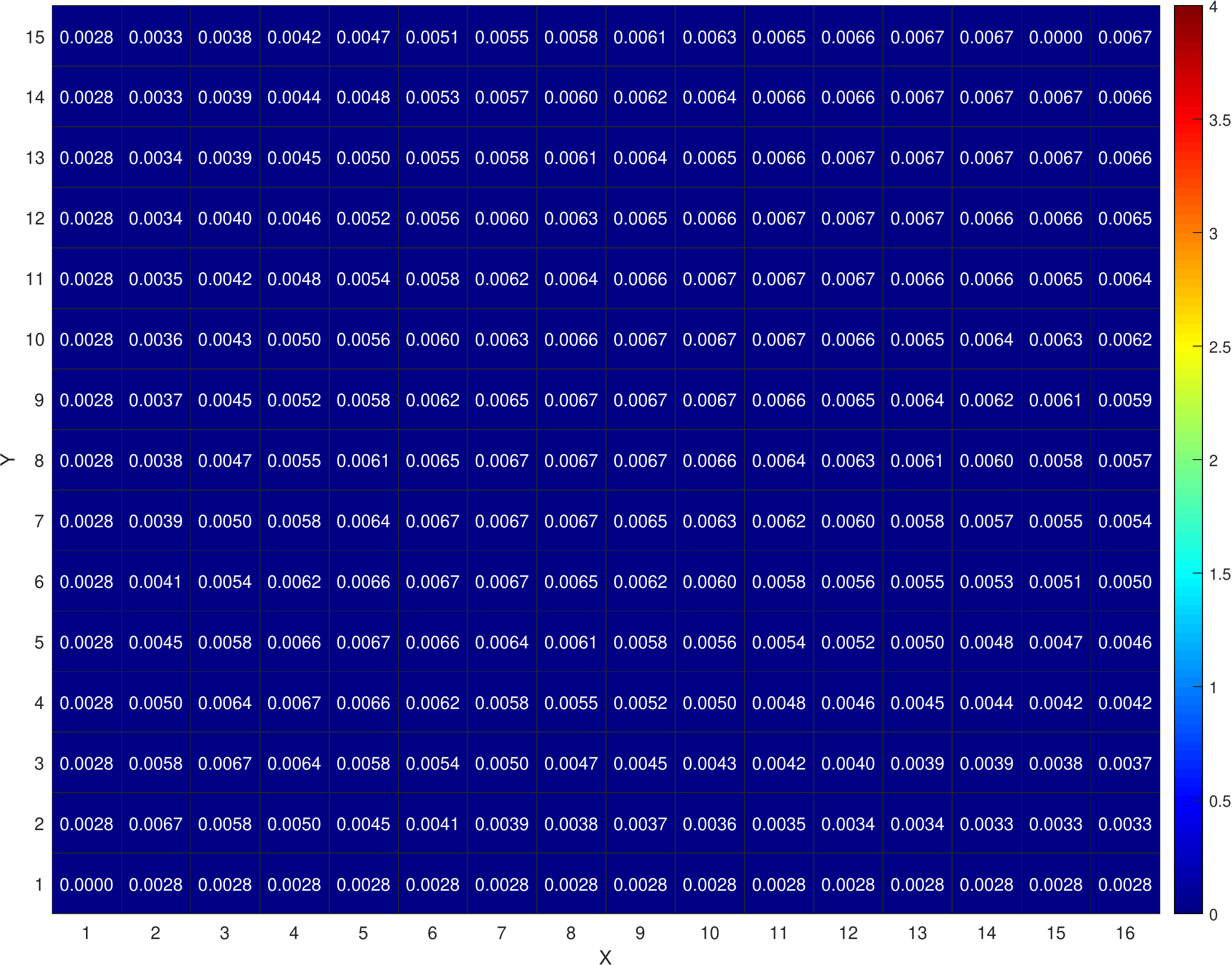}
\caption{SAR reduced with corresponding transmitted power}
\label{fig_sar_after}
\vspace{0.2 in}
\centering
\includegraphics[width = \linewidth]{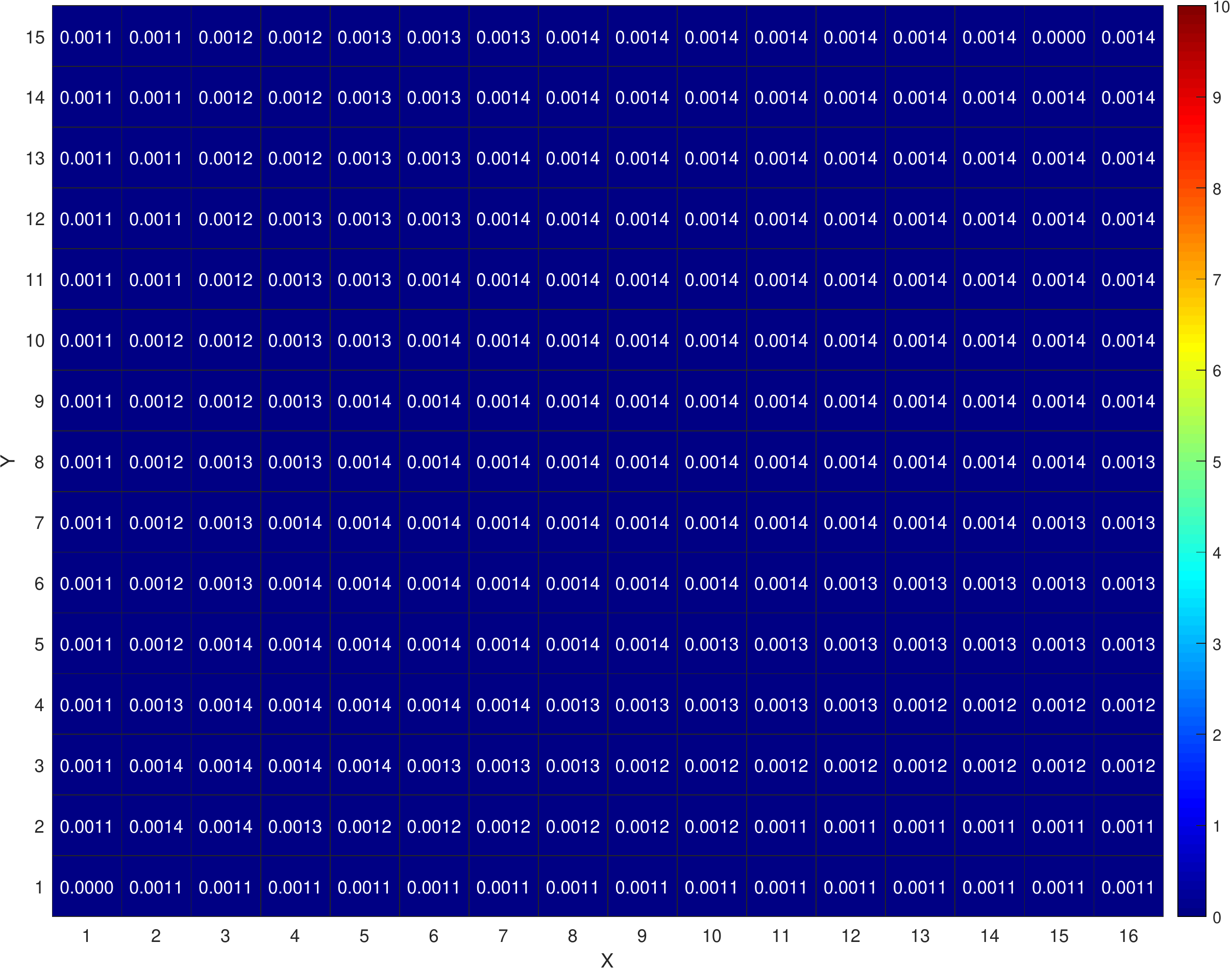}
\caption{PD reduced with corresponding transmitted power in zoom view}
\label{fig_pd_after}
\end{figure}

Figs. \ref{fig_sar_after} and \ref{fig_pd_after} show the heatmap of SAR and total power density using the proposed protocol to reduce the RF EMF in wearable communications. Here a control scheme with reduced transmit power (almost 43$\%$) has been used to make the data rates same using the relay in multi-hop to that of the data rate without the relay being used in single-hop communication.

\subsubsection{Variation of Tx and Rx Positions}
\begin{figure}[t]
\centering
\includegraphics[width = \linewidth]{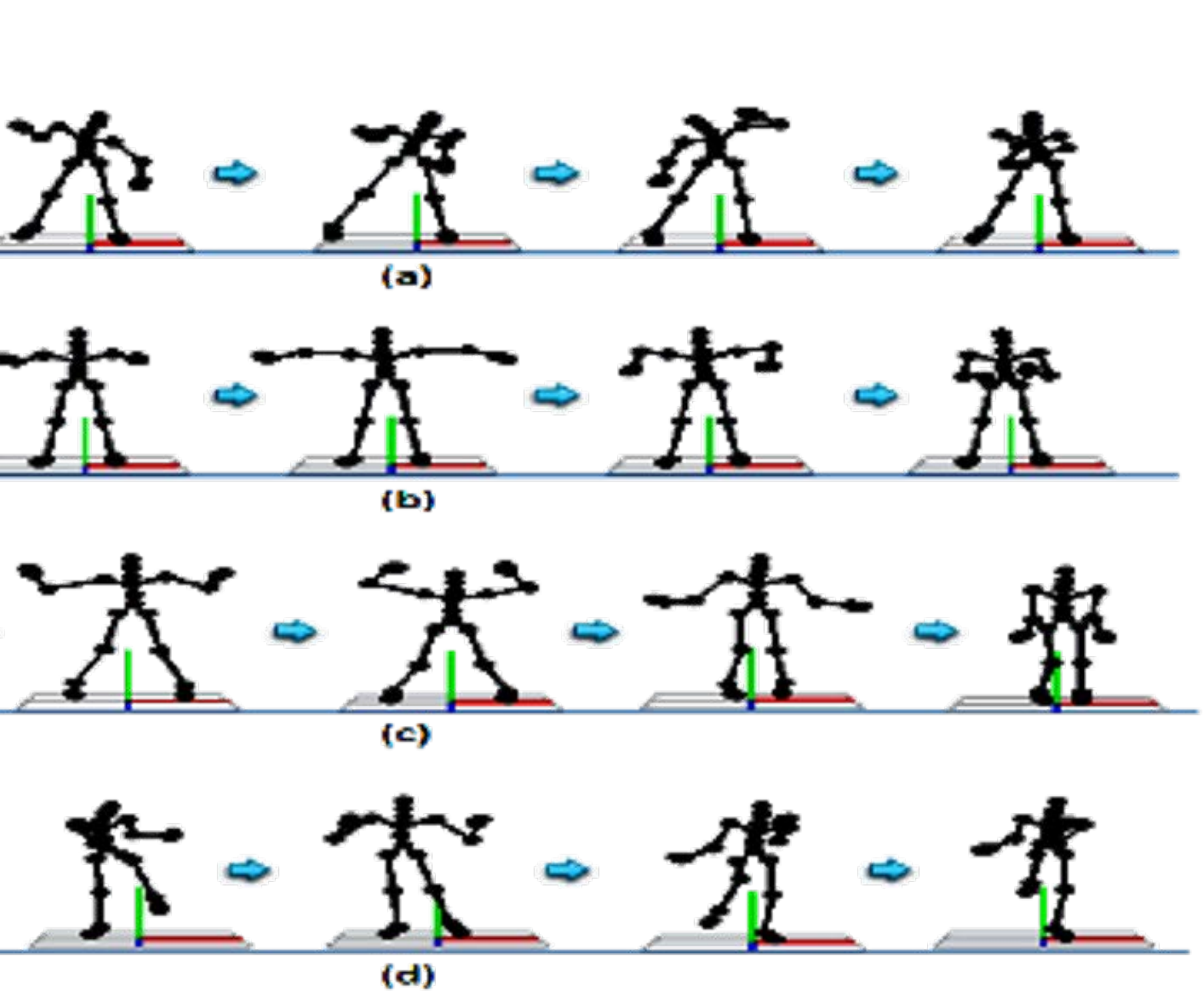}
\caption{Designated exercises moving the Tx and Rx in different orientations \cite{linn}}
\label{fig_exercises}
\end{figure}

Measurements were made taking into account the various body configurations and orientation of the Tx-Rx pair. As the human subject is assumed to perform some designated and cyclic exercises, these devices will move along with the designated trajectories. Fig. \ref{fig_exercises} shows designated exercises moving the Tx and the Rx in different orientations. So the locations of Tx and Rx were varied as well. Different scenarios were tested in this simulation.

\begin{figure}[t]
\centering
\includegraphics[width = \linewidth]{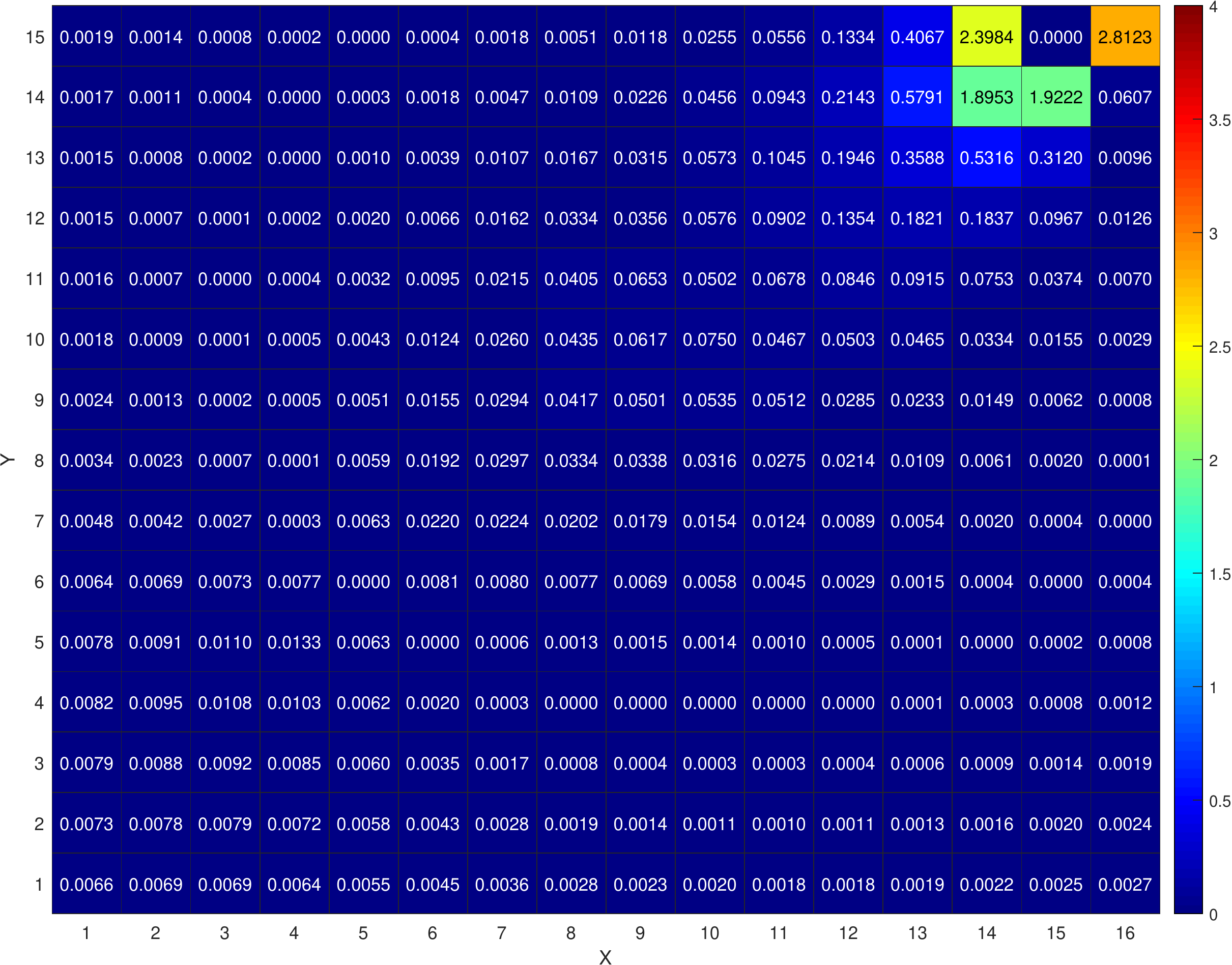}
\caption{SAR varying the Tx locations without Tx power reduction in zoom view}
\label{fig_sar_pos1}
\vspace{0.2 in}
\centering
\includegraphics[width = \linewidth]{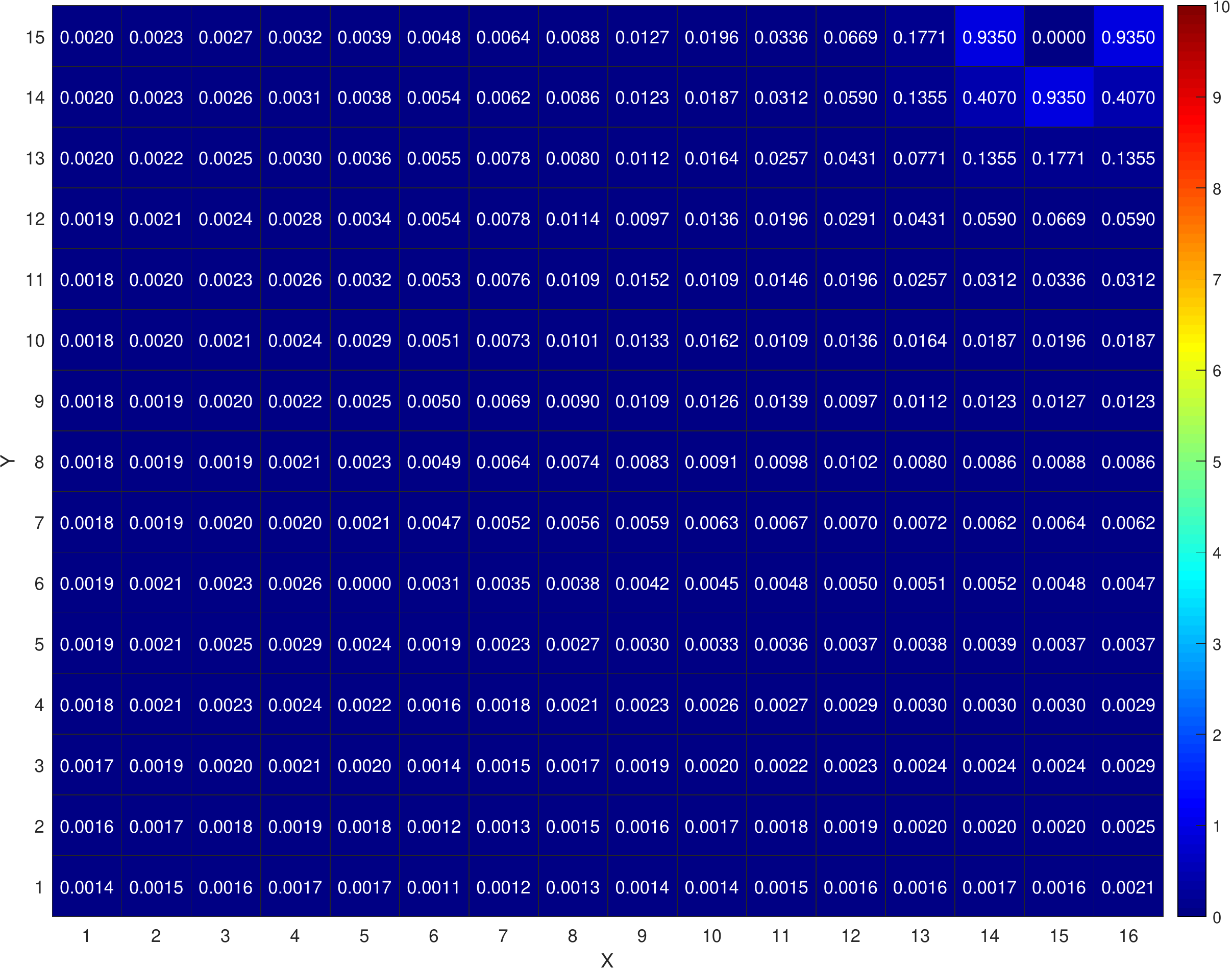}
\caption{PD varying the Tx locations without Tx power reduction in zoom view}
\label{fig_pd_pos1}
\end{figure}

First, the relay (5,6) and the Rx (15,15) were fixed now and the Tx location was varied accordingly in each possible coordinate to communicate with the Rx directly or via the relay communication. Since the data rates do not differ with and without relay, the ``reduced Tx power'' mechanism (see Fig. \ref{fig_flowchart}) is not being used here and the SAR and PD meets the guideline according to the FCC. The scenarios are shown in Figs. \ref{fig_sar_pos1} and \ref{fig_pd_pos1}.

\begin{figure}[t]
\centering
\includegraphics[width = \linewidth]{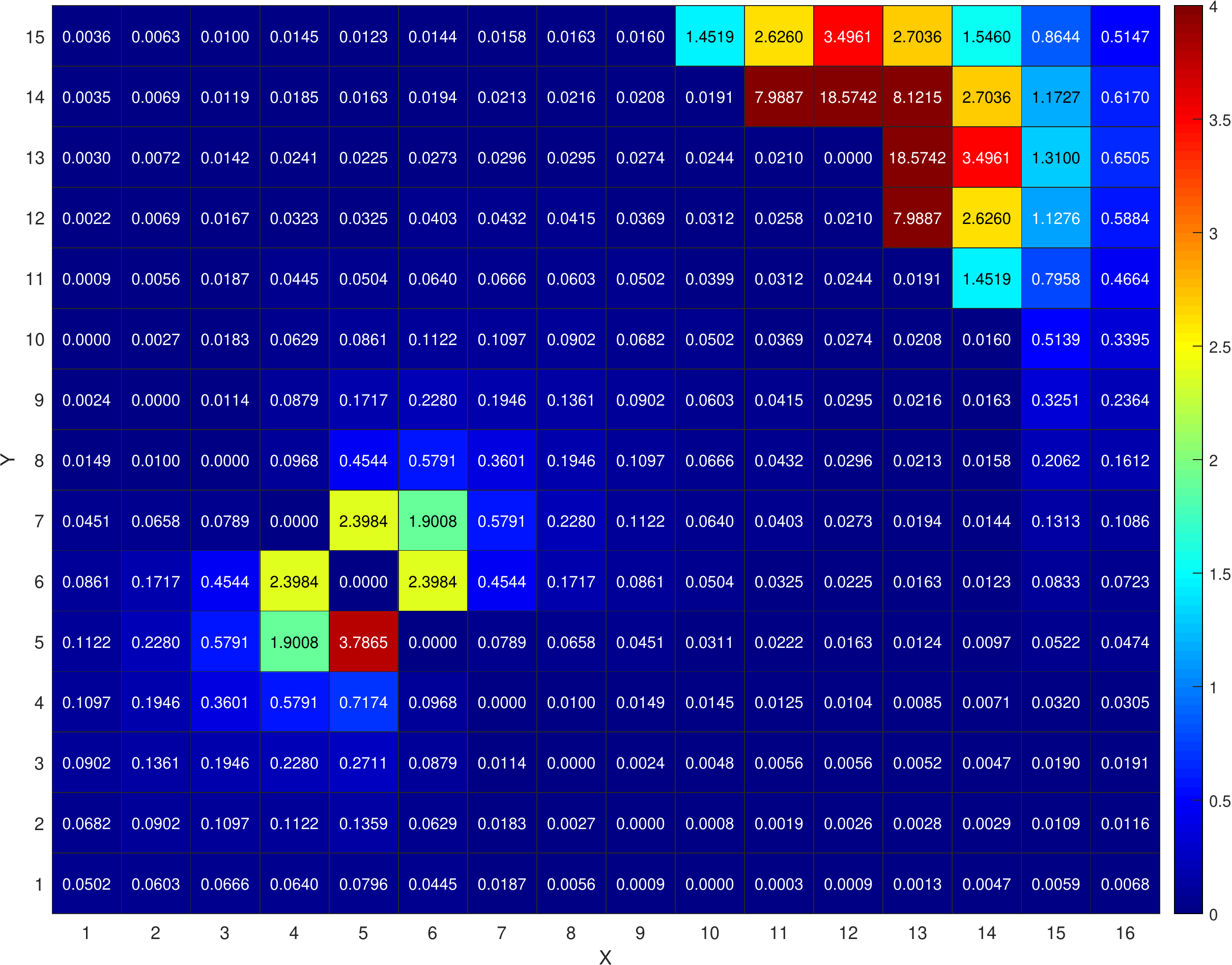}
\caption{SAR varying the Rx locations without Tx power reduction in zoom view}
\label{fig_sar_pos2}
\vspace{0.2 in}
\centering
\includegraphics[width = \linewidth]{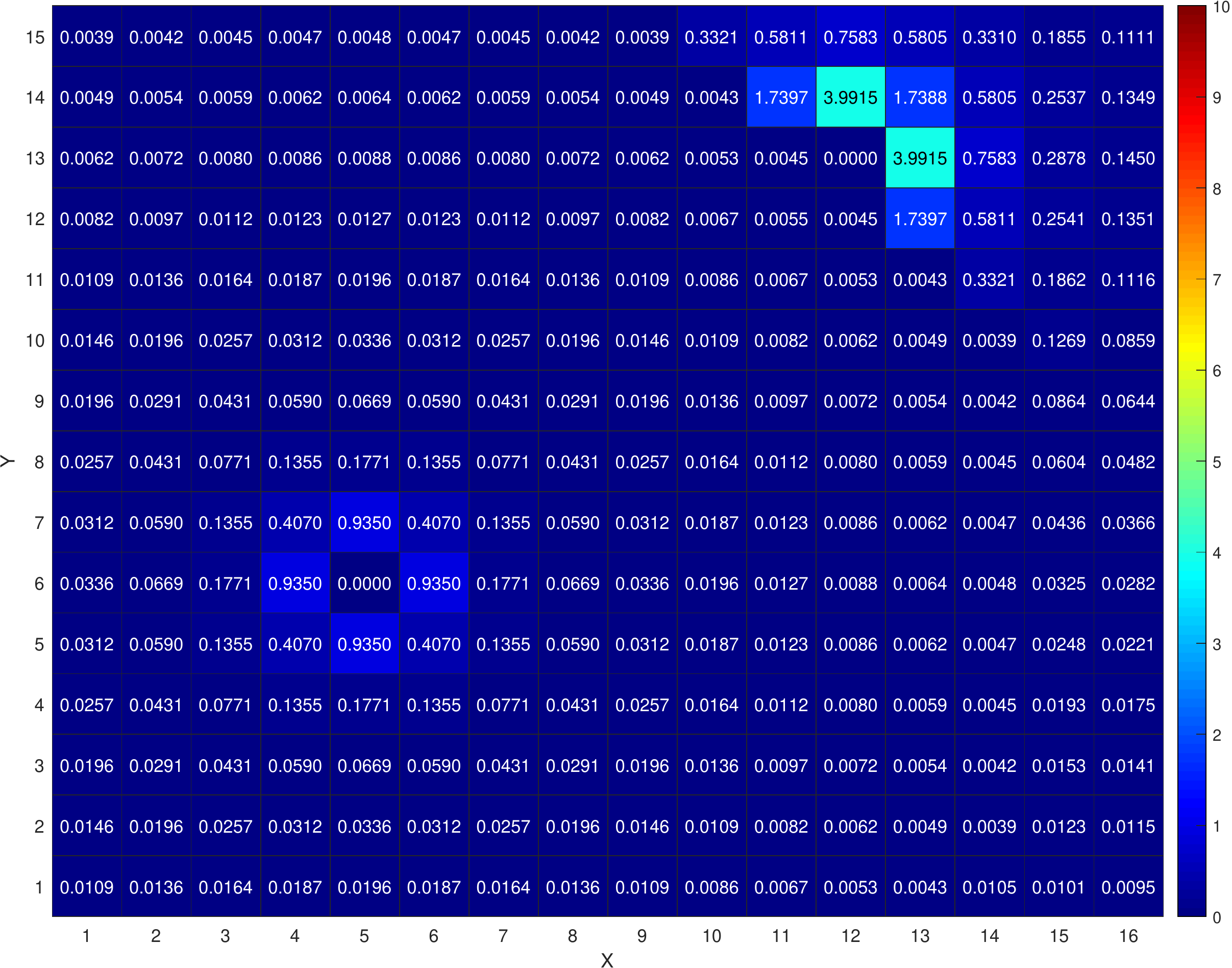}
\caption{PD varying the Rx locations without Tx power reduction in zoom view}
\label{fig_pd_pos2}
\end{figure}

\begin{figure}[t]
\centering
\includegraphics[width = \linewidth]{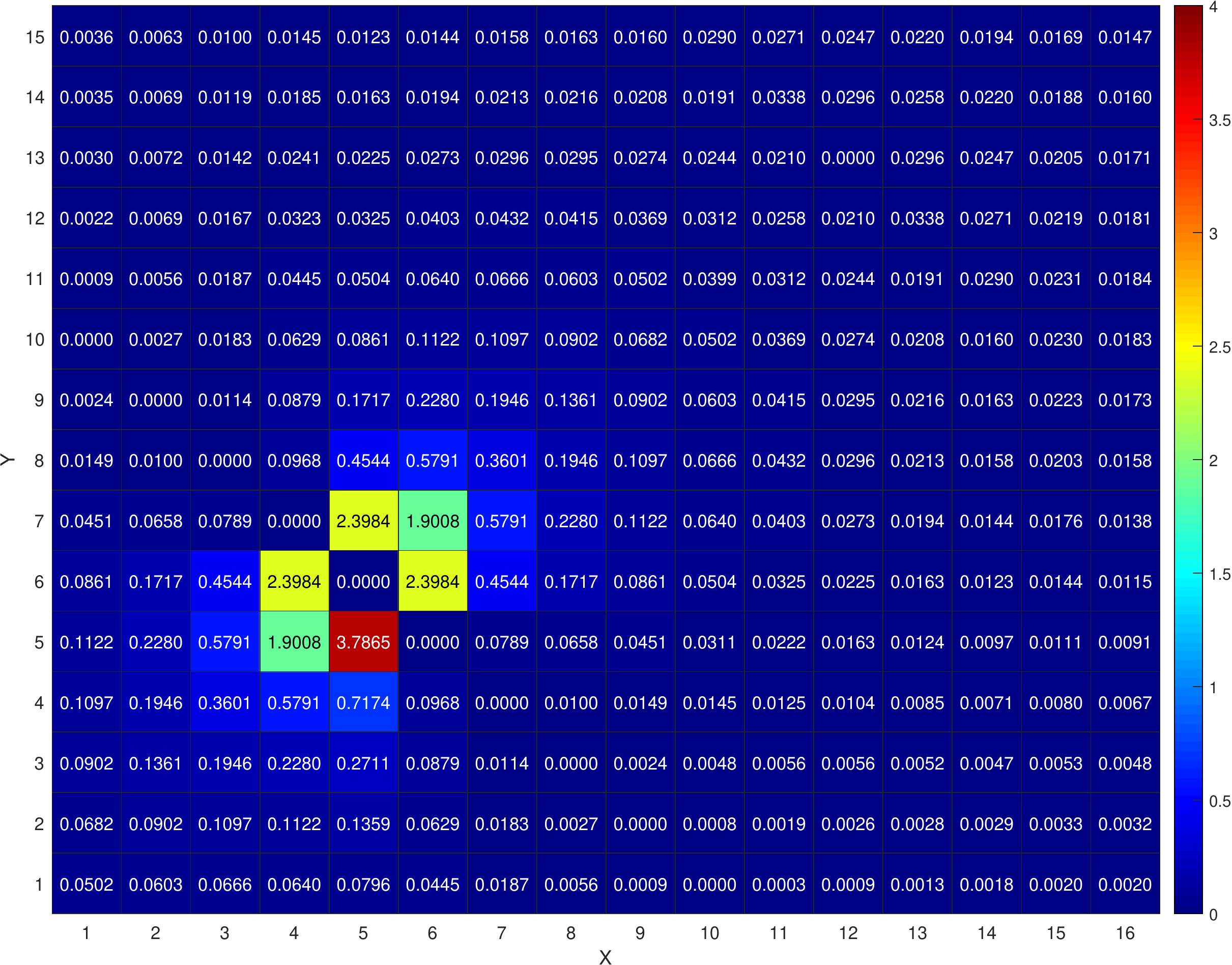}
\caption{SAR varying the Rx locations with Tx power reduction in zoom view}
\label{fig_sar_pos3}
\vspace{0.2 in}
\centering
\includegraphics[width = \linewidth]{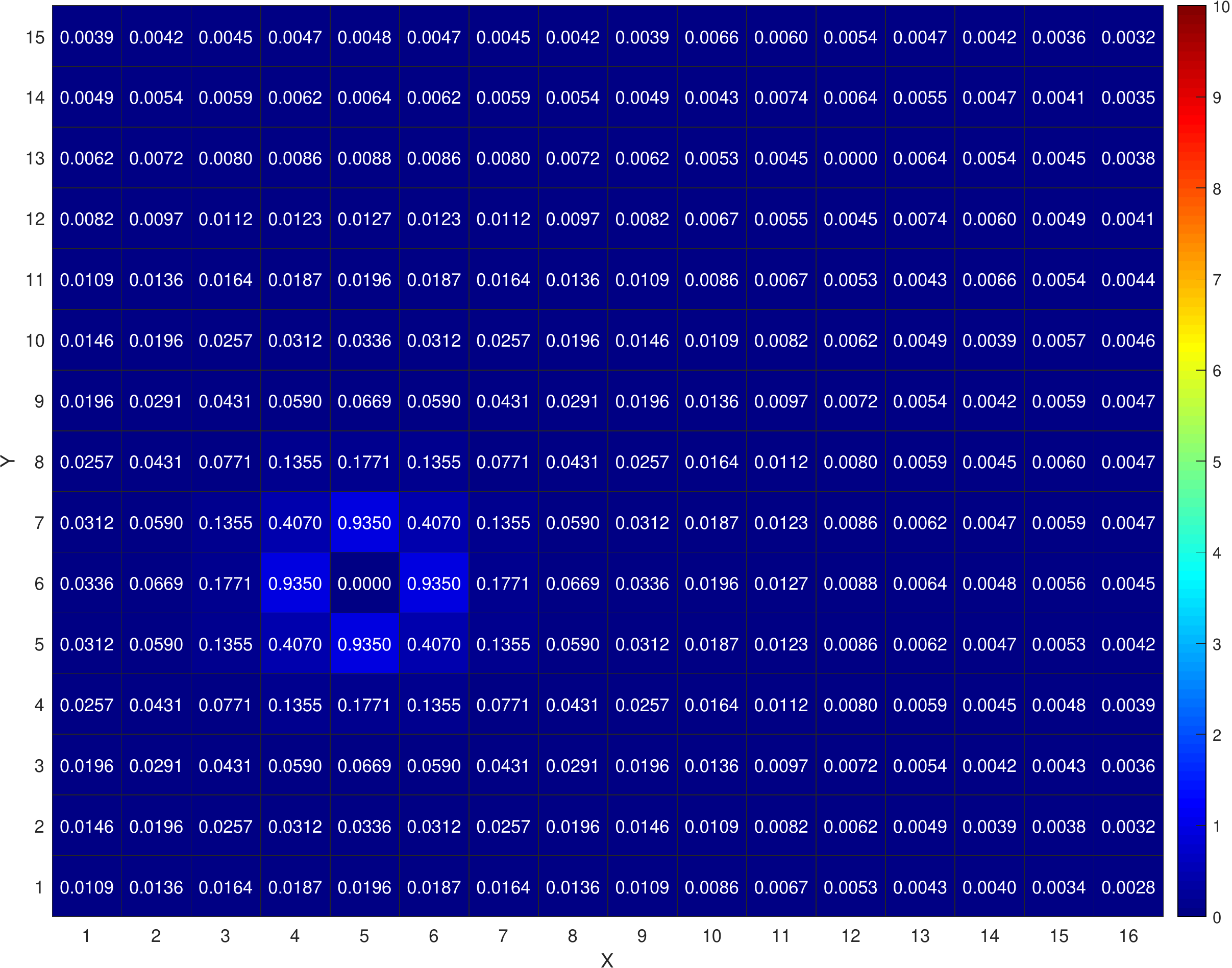}
\caption{PD varying the Rx locations with Tx power reduction in zoom view}
\label{fig_pd_pos3}
\end{figure}

Second, it was tried with varying the position of Rx in every possible location fixing the coordinates of the relay (12,13) and the Tx (5,6) in surface grid. The Tx thus communicates with the Rx directly or via relay communication based on the distance and the heatmaps of SAR and PD plots are shown in Figs. \ref{fig_sar_pos2} and \ref{fig_pd_pos2} accordingly.

After using the mitigation mechanism deploying reduced Pt the SAR and PD plots are as below shown in Figs. \ref{fig_sar_pos3} and \ref{fig_pd_pos3} meeting the regulations based on FCC.
The simulations were carried out on numerical human models and powers of transmission were obtained which ensured less bit error rate. The findings showed that by using these transmitting powers for the system, the SAR value did not surpass the safety guidelines.

To minimize the impact of SAR on human health during transmission, a MAC protocol that tracks and forecasts channel variability and transmission schedules when the RSS is likely to be higher in an opportunistic way is also proposed \cite{banmac}.

\section{Conclusions}
This paper has provided a PD control mechanism for wearable communication devices on RF exposure distinguishing from the prior works, where adverse impacts of wearable communications were investigated in terms of SAR according to the carrier frequency. Some of the devices are found to cause SAR levels that exceed the guidelines. Based on the findings, this paper suggests not to wear the devices on the body continuously for a long time and proposes a new protocol for SAR reduction along with multi-hop communication. The result shows comparison between direct and two-hop wireless communication. Direct communication or single-hop transmission is possible but it impacts the body more with high transmit power. A new protocol has been proposed for SAR and PD reduction as well as RF mitigation for wearable communication. Furthermore, The results show that the proposed protocol for two-hop communication can be more effective in the reduction of SAR. For future work, Simplicity and robustness, can be used to improve packet success rate and different metrics other than PD and SAR can be considered for mitigation of RF exposure. Also there are challenges and enabling technologies going on for wearable communications in 5G mmW. Analysis of RF exposure and mitigation scheme must be considered in such high frequencies as well.


\end{document}